\documentclass[twocolumn,showpacks,aip,jcp,reprint]{revtex4-1}

\usepackage{graphicx}
\usepackage{amssymb}
\usepackage{amsmath,amsfonts,epsfig}
\usepackage[english]{babel}
\usepackage{graphicx}
\usepackage{url}
\usepackage{bm}
\usepackage{subfigure}
\usepackage{color}

\DeclareUrlCommand{\blackurl}{}

\newcommand{\rvec}{\textbf{r}}
\newcommand{\pvec}{\textbf{p}}

\newcommand{\lb}{\lambda_{\rm{B}}}

\newcommand{\dif}{\mathrm{d}}
\newcommand{\V}{V}
\newcommand{\Vion}{V_{\mathrm{out}}}
\newcommand{\Vcol}{V_{\mathrm{in}}}

\newcommand{\iondiam}{d}
\newcommand{\Rijvec}{\textbf{R}_{ij}}
\newcommand{\svec}{\textbf{s}}
\newcommand{\nvec}{\hat{\textbf{n}}}
\newcommand{\Vmmij}{\mathcal{V}^{\mathrm{MM}}_{ij}}
\newcommand{\Vij}{\mathcal{V}_{ij}}
\newcommand{\Vmdij}{\mathcal{V}^{\mathrm{MD}}_{ij}}
\newcommand{\Vddij}{\mathcal{V}^{\mathrm{DD}}_{ij}}
\newcommand{\cref}[1]{(\ref{#1})}
\newcommand{\citereg}[1]{\onlinecite{#1}}

\begin{document}

\title{Electrostatic Interactions between Janus Particles}

\author{Joost de Graaf}%
 \email{j.degraaf1@uu.nl}
\affiliation{%
Soft Condensed Matter, Debye Institute for Nanomaterials Science, Utrecht University,\protect\\ Princetonplein 5, 3584 CC Utrecht, The Netherlands
}%

\author{Niels Boon}%
\affiliation{%
Institute for Theoretical Physics, Utrecht
University, \\ Leuvenlaan 4, 3584 CE Utrecht, The Netherlands
}%

\author{Marjolein Dijkstra}%
\affiliation{%
Soft Condensed Matter, Debye Institute for Nanomaterials Science, Utrecht University,\protect\\ Princetonplein 5, 3584 CC Utrecht, The Netherlands
}%

\author{Ren\'e van Roij}%
 \email{r.vanroij@uu.nl}
\affiliation{%
Institute for Theoretical Physics, Utrecht
University, \\ Leuvenlaan 4, 3584 CE Utrecht, The Netherlands
}%

\date{\today}

\begin{abstract}
In this paper we study the electrostatic properties of `Janus' spheres with unequal charge densities on both hemispheres. We introduce a method to compare primitive-model Monte Carlo simulations of the ionic double layer with predictions of (mean-field) nonlinear Poisson-Boltzmann theory. We also derive practical DLVO-like expressions that describe the Janus-particle pair interactions by mean-field theory. Using a large set of parameters, we are able to probe the range of validity of the Poisson-Boltzmann approximation, and thus of DLVO-like theories, for such particles. For homogeneously charged spheres this range corresponds well to the range that was predicted by field-theoretical studies of homogeneously charged flat surfaces. Moreover, we find similar ranges for colloids with a Janus-type charge distribution. The techniques and parameters we introduce show promise for future studies of an even wider class of charged-patterned particles.
\end{abstract}

\pacs{82.70.Dd, 87.16.D-, 61.20.Qg, 41.20.-q}

\maketitle

\section{\label{EWsec:intro}Introduction}

Electrostatic interactions in suspensions of charged colloids are of paramount importance to the structure and phase behaviour of such systems~\cite{Grier,BelloniR,Hansen,yethiraj_colloidal_2003,Hoffmann,leunissen_ionic_2005,leuniss,Levin6,Palberg0,Palberg1,Bianchi,Vissers}. The Derjaguin Landau Verwey Overbeek (DLVO)~\cite{Derjaguin,Verwey} theory is the most well-known theory by which the pair interactions between screened charged particles can be described. However, DLVO theory only describes the interactions between homogeneously charged spherical particles and is therefore a \emph{monopole} theory. The rapidly-growing zoo~\cite{Glotzer,Pine} of new colloidal particles demands similar theories that are equipped to deal with anisotropy in shape and size.

For classical, spherically charged, particles, DLVO theory has proven to be a powerful means to describe the electrostatic properties of systems; mostly for high-polarity solvents, low surface charge, and high ionic strength. The nonlinear Poisson-Boltzmann (PB) approach~\cite{Gouy,Chapman,Alexander} extends this range, although analytic results are only possible for a limited number of systems. PB theory is based on a mean-field approximation that ignores ion-ion correlations, which is justified only for sufficiently high temperatures, not too apolar media, and monovalent ions at reasonable concentrations. In regimes where ion-ion correlations are important Strong-Coupling (SC) theory~\cite{Rouzina,Moreira,Grosberg} may be applied. Several modifications to PB theory exist~\cite{Linse8,Trizac1,Trizac3,Yu_wu_gao,Levin2,Yaakov,Lue0,Lue1,Levin6}, including modifications of the traditional PB theory that account for finite-size ions.

In the computational field, both Monte Carlo (MC) and Molecular Dynamics (MD) techniques have been developed to analyse charged particles suspended in an electrolyte. In \emph{primitive-model} simulations, ions are taken into account explicitly, whilst the solvent is modelled as a dielectric continuum~\cite{Linse3,guldbrand,Degreve,orkoulas,Valeriani0,Valeriani1}. Ewald Sums~\cite{Ewald} usually provide the basis for calculating the long-ranged Coulomb contribution to the total energy of a system with periodic boundary conditions. However, employing Ewald Sums is computationally expensive and the systems that can be studied in the primitive model are therefore typically small. When studying charged colloids suspended in an electrolyte it is desirable to coarse-grain the system and use the much shorter ranged effective interactions between the particles instead of accounting for the ions explicitly. Most simulation studies therefore consider systems where the interactions between the colloids can be modelled by a DLVO pair potential~\cite{Stradner,Sanz,Toledano,Vissers,colombo}. We note that faster Ewald-Sums implementations exist, e.g., mesh-based approaches~\cite{deserno0,deserno1}, but these cannot outperform a course-grained method that does not take the ions into account at all in a system of charged colloidal particles.

In this paper we investigate the range of applicability for nonlinear PB theory to accurately describe the behaviour of the ion density around charged \emph{heterogeneous} particles. This allows us to quantify the parameter regime for which a (multipole-expanded) DLVO approximation may be applied to describe pair interactions in coarse-grained simulations, since (the electrostatic part of) such DLVO theories can be derived using PB approximations. Charge-patterned particles have already been studied using the DLVO approximation by partitioning the surface charge over a finite number of point-Yukawa charges with different prefactors~\cite{Hong} to obtain effective pair interactions. For charge-patterned particles this approximation still generally results in an expensive calculation of the pair interaction as a function of the separation and orientation. Moreover, the point-Yukawa description inadequately accounts for the hard core of the particle, i.e., it implicitly assumes that ions can penetrate the colloid~\cite{Boon}. In this paper a correct, simple, and therefore computationally far more efficient DLVO-multipole approximation is derived for the effective interaction between two charge-patterned particles. The multipole-based effective interactions have an enormous potential for use in simulation studies to explore the phase behaviour of previously inaccessible systems.

For this study we focus on particles with a Janus-type charge pattern. The term \emph{Janus} refers to the two-faced Roman god of doors and was introduced to describe colloid properties in 1988~\cite{Casagrande}. A Janus particle~\cite{Casagrande,Gennes,Walther} consists of two opposing parts (usually hemispheres) with different properties for the wetting, charge, chemical functionality, etc. The past decade has seen a marked increase in the ability to synthesize such Janus colloids~\cite{roh_biphasic_2005,Janus,JB3,Schultz,Grafting,Tripathy,LargeSC} and their use in self-assembly experiments. Many interesting structures have been found~\cite{Hong,chen_directed_2011} and questions have been raised on how to approach simulations of such systems. With our study we aim to address some of these questions for charged Janus particles in an electrolyte, in much the same way as the pioneering simulation studies that probed the applicability of the common DLVO/PB approximation for homogeneously charged particles~\cite{Linse3,Degreve}.

In Section~\ref{EWsec:method} we introduce the methods by which we compute the ion density around charged Janus particles: primitive-model MC simulations (\ref{EWsub:simulation}) and nonlinear PB theory (\ref{EWsub:theory}). We discuss the results of our investigation in Section~\ref{EWsec:result}, which is divided into four parts. In Section~\ref{EWsub:compare} we introduce the method, based on Fourier-Legendre (FL) mode decomposition, by which we compare the MC and PB results. This method is applied for a homogeneously charged particle in Section~\ref{EWsub:homogen}, where we also investigate the relation to the field-theoretical results of Refs.~\citereg{Boroud,Punk} for homogeneously charged flat surfaces. In Section~\ref{EWsub:janus} we extend our results to a Janus dipole and show that there is remarkable correspondence with the results for a homogeneously charged sphere. We consider a particle with a single charged hemisphere in Section~\ref{EWsub:hemi}. Throughout these sections we give explicit recipes for calculating the pair interactions between such particles within the PB approximation that we are testing. The interested reader is referred to the Appendix for a derivation of these pair interactions. We discuss our findings, comment on the potential synergy between simulation methods and theoretical results, and present an outlook in Section~\ref{EWsec:disc}.

\section{\label{EWsec:method}Simulations and Theory}

In the following we consider a system of spherical charge-patterned colloids with radius $a$ suspended in an electrolyte. The colloid volume fraction is denoted by $\eta$. We studied three types of charge distribution for the colloids. (i) A homogeneous surface charge of $Ze$, with $Z > 0$ the number of charges and $e$ the elementary charge. (ii) A perfectly antisymmetric surface charge, with charges $Ze/2$ and $-Ze/2$ homogeneously distributed over the particle's upper and lower hemisphere, respectively. (iii) A homogeneously charged upper hemisphere with charge $Ze$ and a completely uncharged lower hemisphere. Unless stated otherwise $Z = 100$ throughout this paper. We assume that there is a perfect dielectric match between the colloid, the ions, and the medium to avoid any dielectric boundary effects.

\subsection{\label{EWsub:simulation}Ewald Sums and Monte Carlo Simulations}

To study the systems described above by MC simulations we turn to the primitive model, for which the ions are represented by charged spheres and the solvent is treated as a dielectric continuum. To simplify the calculations we study only one of these particles, which we locate at the centre of a volume $V_{\mathrm{cell}} = 4 \pi a^{3}/(3 \eta)$. We apply periodic boundary conditions to this volume to account for the fact that we are in principle interested in a system which contains many colloids. The particle's (heterogeneous) surface charge is specified by $100$ charge sites distributed over this surface, which can be positively or negatively charged, or which do not have charge. These charge sites on the colloid are chosen according to the optimal packing of 100 points on a sphere~\cite{SiteDistr} to ensure that they are spaced as homogeneously as possible.

The number of free monovalent ions $N$ in the volume $V_{\mathrm{cell}}$ is fixed, i.e., we are interested in an average ion concentration $N/V_{\mathrm{cell}}$ for the system that we approximate by our one-colloid calculation. We only consider systems for which a monovalent salt has been added to enhance the screening effected by the counter ions to the particle's charge. The balance between the number of positive $N_{+}$ and negative $N_{-}$ ions ($N = N_{+} + N_{-}$) is such that the volume, and thereby the entire system, is charge neutral. For the monopole and charged hemisphere we require $Z + N_{+} - N_{-} = 0$ and for the Janus dipole $N_{+} = N_{-}$.

To sample phase space Monte Carlo (MC) simulations are performed in the isothermal-isochoric (canonical, $NVT$) ensemble. We consider a cubic simulation box of length $L = 50d$ ($\eta = (4\pi/3)(a/L)^{3}$, $V_{\mathrm{cell}} = L^3$), for which we employ periodic boundary conditions. Here $d$ is the ion diameter and we assume all ions to be the same size. The spherical colloidal particle is located at the centre of the box (the origin). The particle's rotational symmetry axis is chosen parallel to one of the box' ribs for the Janus-type charge distributions.

The ion-ion pair potential is a combination of a Coulomb and a hard-core interaction part:
\begin{equation}
\label{EWeq:ion_inter} \mathcal{U}_{\mathrm{II}}(\mathbf{r}_{i},\mathbf{r}_{j}) = \frac{q_{i}q_{j}e^{2}}{4 \pi \epsilon_{0} \epsilon} \frac{1}{ \vert \mathbf{r}_{i} - \mathbf{r}_{j} \vert } + \left\{ \begin{array}{cl} \infty, & \vert \mathbf{r}_{i} - \mathbf{r}_{j} \vert \le d ; \\ 0,  & \vert \mathbf{r}_{i} - \mathbf{r}_{j} \vert > d , \end{array} \right.
\end{equation}
with $\mathbf{r}_{i}$ and $\mathbf{r}_{j}$ the position of ions $i$ and $j$ with respect to the colloid's centre, respectively. The function $\vert \cdot \vert$ gives the Euclidean norm of a vector, $q_{i} = \pm 1$ the sign of the $i$-th ion's charge, $\epsilon_{0}$ the permittivity of vacuum, and $\epsilon$ the relative dielectric constant of the medium, ions, and particle. The interaction between a charge $q_{i}e$ on the particle located at $\mathbf{r}_{i}$, with $\vert\mathbf{r}_{i} \vert = a - d/2$, and an ion with charge $q_{j}e$ located at $\mathbf{r}_{j}$ is given by
\begin{equation}
\label{EWeq:mi_inter} \mathcal{U}_{\mathrm{SI}}(\mathbf{r}_{i},\mathbf{r}_{j}) = \frac{q_{i}q_{j}e^{2}}{4 \pi \epsilon_{0} \epsilon} \frac{1}{ \vert \mathbf{r}_{i} - \mathbf{r}_{j} \vert } + \left\{ \begin{array}{cl} \infty, & \vert \mathbf{r}_{j} \vert \le a+d/2 ; \\ 0, & \vert \mathbf{r}_{j} \vert > a+d/2 . \end{array} \right. 
\end{equation}
The coupling between periodicity and the (long-range) Coulomb interactions, Eqs.~\cref{EWeq:ion_inter} and~\cref{EWeq:mi_inter}, is taken into account using Ewald Sums with conductive boundary conditions~\cite{Ewald,Frenkel_Smit}. The total electrostatic energy $U_{\mathrm{C}}$ of a particular configuration may be written as
\begin{flalign}
\nonumber & \frac{4 \pi \epsilon_{0} \epsilon}{e^{2}}U_{\mathrm{C}} = \\
\nonumber & \frac{1}{2L^{3}} \sum_{\mathbf{k} \ne \mathbf{0}} \left[ \frac{4 \pi}{\vert \mathbf{k} \vert^{2}} \left\vert \sum_{j=1}^{\tilde{N}} q_{j} \exp \left( i \mathbf{k} \cdot \mathbf{r}_{j} \right) \right\vert^{2} \exp \left( -\frac{\vert \mathbf{k} \vert^{2}}{4 \gamma} \right) \right]  \\
\label{EWeq:ewaldsums} & -\sqrt{ \frac{\gamma}{\pi} } \sum_{i=1}^{\tilde{N}} q_{i}^{2} + \frac{1}{2} \sum_{i\ne j}^{\tilde{N}} \frac{q_{i} q_{j} \mathrm{erfc} \left( \sqrt{\gamma} \vert \mathbf{r}_{i} - \mathbf{r}_{j} \vert \right)}{\vert \mathbf{r}_{i} - \mathbf{r}_{j} \vert} ,
\end{flalign}
where summation is over both the ions and the charge sites, i.e., $\tilde{N}$ is the total number of charges in the system, both free and fixed; $\mathbf{k} \equiv (2 \pi / L ) \mathbf{l}$ is a Fourier space vector, with $\mathbf{l} \in \mathbb{Z}^{3}$; $\gamma$ is the Ewald convergence parameter~\cite{Frenkel_Smit}; and $\mathrm{erfc}(\cdot)$ is the complementary error function. One can safely ignore the site-site interactions in Eq.~\cref{EWeq:ewaldsums}, because this gives a constant contribution to the electrostatic energy $U_{\mathrm{C}}$. The self-energy term also drops out of the energy difference, on which the acceptance criterion for the MC trial moves is based~\cite{Frenkel_Smit}.

For our simulations we employ the following parameters. (i) Each run consists of 100,000 MC equilibration cycles, where 1 MC cycle is understood to be one trial (translation) move per free ion. (ii) This equilibration is followed by a production run of 250,000 MC cycles to determine the ensemble-averaged ion density profiles $\rho_{\pm}(\mathbf{r})$, with $\mathbf{r}$ the position with respect to the centre of the colloid. (iii) The step size for the ion translational moves is in the range $[0,5d]$ and it is adjusted to yield an acceptance ratio of $0.25$. (iv) For the Ewald Sums the real space cut-off radius for the third term in Eq.~\cref{EWeq:ewaldsums} is set to $L/2.5$, and we use $\gamma = 0.03$ and $\vert \mathbf{l} \vert < 6$. This choice of Ewald parameters gives a reasonable approximation to the value of the electrostatic interaction energy. Doubling and halving the number of cycles for several runs showed that the MC parameters give sufficiently equilibrated results for most systems. A possible exception to the perceived equilibration is deep inside the strong-coupling regime, where ion-ion correlations play an important role, as we will explain in Section~\ref{EWsub:compare}.

\subsection{\label{EWsub:theory}The Poisson-Boltzmann Approach}

The spherical particle of radius $a$ in a cubic box ($L \times L \times L$) models a system with colloid volume fraction $\eta = (4\pi/3)(a/L)^{3}$. The equivalent system in PB theory is described using a spherical Wigner-Seitz (WS) cell model~\cite{Wigner_Seitz,Marcus,Ohtsuki,Alexander}, where the radius of the WS cell is given by
\begin{equation}
\label{EWeq:Rws} R = \left( \frac{3 L^{3}}{4 \pi} \right)^{1/3} = a \eta^{-1/3},
\end{equation}
and the colloid is located at the centre of the cell. The choice of $R$ ensures that the volumes, and therefore the average density of colloids/ions is the same as in the cubic box of the MC simulations. PB theory is applied, in accordance with the procedure outlined in Refs.~\citereg{Eggen,eggen_phd,boon_phd}, to determine the dimensionless electrostatic potential $\phi(\mathbf{r})$ and the associated ion density profiles $\rho_\pm(\mathbf{r})$ around the colloid.

In our MC simulations the hard-core interaction between the ions and the colloid prevent the ions from approaching the colloid's centre closer than a distance of $a + d/2$. We therefore assume the same spherical hard-core exclusion volume for the point ions in PB theory. The colloid's surface charge density is given by $q(\mathbf{r})$, which is only nonzero when $\vert \mathbf{r} \vert = a$; the spatial integral over $q(\mathbf{r})$ gives the total colloid charge. The PB equation for this system may now be written as
\begin{equation}
\label{EWeq:nabla2phi} \nabla^2 \phi(\mathbf{r}) = 4\pi \lambda_{\mathrm{B}} q(\mathbf{r}) + \left\{ \begin{array}{cl} 0 & \vert \mathbf{r} \vert \leq a+d/2 \\ \kappa^2 \sinh ( \phi(\mathbf{r} ) ) & \vert \mathbf{r} \vert > a+d/2 \end{array} \right. ,
\end{equation}
where $\kappa^2 = 8\pi\lambda_{\mathrm{B}}\rho_{s}$ (such that $\kappa^{-1}$ is the Debye screening length), with $\rho_{s}$ the (yet unknown) bulk ion density and $\lambda_{\mathrm{B}}$ the Bjerrum length
\begin{equation}
\label{EWeq:bjerrum} \lambda_{\mathrm{B}} = \frac{e^{2}}{4 \pi \epsilon_{0} \epsilon k_{\mathrm{B}} T},
\end{equation}
with $k_{\mathrm{B}}$ Boltzmann's constant and $T$ the temperature. We impose the following boundary condition
\begin{equation}
\label{EWeq:boundary} \left.\nabla \phi(\mathbf{r}) \cdot \bm{\hat{r}}\right|_{r=R} = 0 ,
\end{equation}
with $\bm{\hat{r}} \equiv \mathbf{r}/\vert \mathbf{r} \vert$, on the edge of the spherical cell to ensure that the normal component of the electric field vanishes at the boundary, i.e., the WS cell is charge neutral.

To solve Eq.~\cref{EWeq:nabla2phi} with the above boundary conditions the charge density $q(\mathbf{r})$ and the electrostatic potential $\phi(\mathbf{r})$ are expanded into a complete set of Legendre polynomials as
\begin{eqnarray}
\label{EWeq:qexpand} q(\mathbf{r}) & = &\sum_{\ell} \sigma_{\ell} \delta(r-a) P_{\ell} (x); \\
\label{EWeq:phiexpand} \phi(\mathbf{r}) & = & \sum_{\ell} \phi_{\ell}(r) P_{\ell} (x),
\end{eqnarray}
with $\sigma_{\ell}$ and $\phi_{\ell}(r)$ the surface-charge and potential modes, respectively. Here $r \equiv \vert \mathbf{r} \vert$ and $x \equiv \mathbf{r} \cdot \bm{\hat{z}}$, with $\bm{\hat{z}}$ the orientation of the colloid's rotational symmetry axis. The $P_{\ell}(\cdot)$ are $\ell$-th order Legendre polynomials, i.e.,
\begin{eqnarray}
\label{EWeq:LP0} P_{0}(x) & = & 1; \\
\label{EWeq:LP1} P_{1}(x) & = & x; \\
\label{EWeq:LP2} P_{2}(x) & = & \frac{1}{2}\left( 3 x^{2} - 1 \right); \\
\label{EWeq:LP3} P_{3}(x) & = & \frac{1}{2}\left( 5 x^{3} - 3 x \right); \\
\nonumber & \vdots & \\
\label{EWeq:LPl} P_{\ell}(x) & = & 2^{\ell} \sum_{k = 0}^{\ell} x^{k} \binom{\ell}{k} \binom{\frac{\ell + k - 1}{2}}{\ell},
\end{eqnarray}
where the expressions between brackets in Eq.~\cref{EWeq:LPl} denote binomial coefficients. The nonlinear PB equation [Eq.~\cref{EWeq:nabla2phi}] is likewise expanded using Fourier-Legendre (FL) mode decomposition and Taylor expansion of the $\sinh (\cdot)$ term around the monopole potential $\phi_{0}(r)$. The higher-order expansion coefficients contain products of Legendre polynomials which effects couplings between the various modes. These products must be rewritten as a sum of single Legendre polynomials~\cite{Legendre_Prod} to solve for the separate modes using an iterative procedure. The mode coupling this induces necessitates the analysis of a significant number of multipoles even if, for example, only the dipole mode is of interest. References~\citereg{Boonpatchy,Boon} can be consulted for more information on the procedure of mode expansion to solve the PB equation for heterogeneously charged colloids.

It is important to note that the PB theory treats the screening ions in the grand-canonical ($\mu V T$) ensemble. The MC simulations were however performed in the canonical ensemble, where the number of ions is fixed, to allow for faster exploration of phase space. We fit the bulk ion concentration $\rho_{s}$ in PB theory to ensure that the number of positive and negative ions in the WS cell corresponds to the number of ions in the MC simulation box. We consider this condition, coupled with the fact that we study the same colloid volume fraction $\eta$ in both approaches, sufficient to justify comparison of the results in the two ensembles. The bulk ion concentration is fitted according to the criterion
\begin{equation}
\label{EWeq:Nws} N_{\pm} = N_{\mathrm{\pm, PB}} \equiv \int \dif \mathbf{r} \rho_{\pm,\mathrm{PB}}(\mathbf{r}),
\end{equation}
where the integration is over the region $\vert \mathbf{r} \vert \in [a + d/2, R]$. One of the two equations is redundant, since solving for $N_{+}$ is equivalent to solving for $N_{-}$. The appropriate bulk ion concentration $\rho_{s}$, which comes into the right-hand side of Eq.~\cref{EWeq:Nws} via the dependence of $\rho_{\pm,\mathrm{PB}}(\mathbf{r}) = \rho_{s}\exp( \mp \phi( \mathbf{r} ) )$ on this concentration, is established using an iterative procedure. All PB results presented in this paper were obtained on an equidistant radial grid of 2,000 points for $\vert \mathbf{r} \vert \in [a + d/2, R]$ by $5$-th order Taylor expansion of $\sinh(\cdot)$ using $6$ multipole modes.

\section{\label{EWsec:result}Ionic Screening of Janus Particles}

In this section we describe our results for the comparison of ion density profiles obtained by MC simulations and by PB theory. A total of $99$ systems are considered for each of the three charge-patterned colloids. We use three particle radii $a = 5d$, $10d$, and $15d$. For every particle radius $a$, three salt concentrations are studied: $125$, $250$, and $375$ monovalent cations and anions, respectively, are added to the counter ions already present in the system. This gives $N_{\pm} = 175$, $300$, and $425$ for the Janus dipole. For the homogeneously and hemispherically charged particle $N_{+} = 125$, $250$, and $375$, when $N_{-} = 225$, $350$, and $475$, respectively. We consider $11$ Bjerrum lengths $\lambda_{\mathrm{B}}/d = 0.01$, $0.05$, $0.1$, $0.25$, $0.5$, $1.0$, $2.0$, $4.0$, $6.0$, $8.0$, and $10.0$ for each $a$ and $N_{\pm}$ combination.

We subdivided the results sections on the basis of the surface-charge pattern studied. For each type of charge pattern, we consider the range where PB theory accurately describes the system. We also give a multipole-expanded DLVO description for the effective pair potential between two such objects (for the Janus dipole and charged hemisphere), which may be used in this range to model the pair interactions in coarse-grained simulations.

\subsection{\label{EWsub:compare}Method of Comparison}

\begin{figure}
\begin{center}
\includegraphics[width= 8cm]{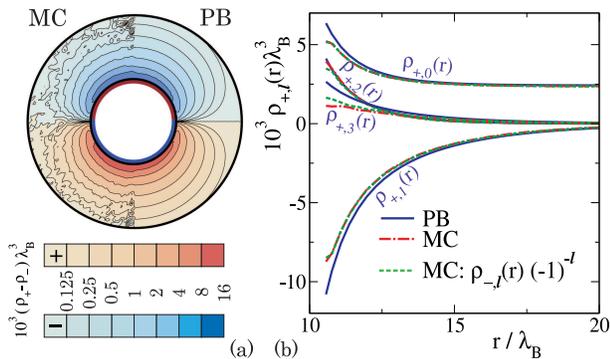}
\end{center}
\caption{\label{EWfig:dipole_moment} A comparison between MC and PB results showing the ion densities around a Janus dipole for $N_{\pm}=300$, $\lb = \iondiam$, and $a=10 \lb$. In (a) the contour plot shows the net charge density $\rho_+(\rvec) - \rho_-(\rvec)$: the MC result (left) and the nonlinear PB solution (right). In (b) the same data is represented using a Fourier-Legendre mode expansion of the charge density $\rho_{\pm,\ell}(r)$, for $\ell = 0$, $1$, $2$, and $3$. Note that the negative modes in (b) can in this case be mapped onto the positive modes by multiplication with $(-1)^{\ell}$.}
\end{figure}

Figure~\ref{EWfig:dipole_moment} shows an example of our results for a typical set of parameters: $a = 10d$, $\lambda_{B} = d$, and 250 added anions and cations, respectively. The azimuthal average of the net ionic charge [$\rho_{+}(\mathbf{r}) - \rho_{-}(\mathbf{r})$] is shown for the Janus dipole (Fig.~\ref{EWfig:dipole_moment}a) to give insight into the shape of the density profiles. The multipole-expanded cation-density profiles (Fig.~\ref{EWfig:dipole_moment}b; $\rho_{+,\ell}(r)$ for $0 \le \ell \le 3$) further illustrate the level of correspondence between the MC and PB result for this system. Due to the antisymmetry of the problem combined with the fact that $P_{\ell} (x) = (-1)^{\ell} P_{\ell} (-x)$ the anion densities follow as $\rho_{-,\ell}(r) = \rho_{+,\ell}(r)(-1)^{\ell}$ for the Janus dipole. The system is in the sufficiently dilute and weak-coupling regime for the ion-ion interactions ($\alpha \approx 0.06$), which explains the good agreement between both methods. Here we use the association-parameter $\alpha \in [0,1]$ introduced in Ref.~\citereg{Valeriani1}, which gives the equilibrium fraction of the available ions in the electrolyte that have formed pairs, to quantify the extent to which strong-coupling effects occur. The definition of $\alpha$ in terms of our variables reads
\begin{eqnarray}
\label{EWeq:alpha} \alpha & = & 1 - \frac{1}{2 K\rho_{s}} \left( \sqrt{1 + 4 K \rho_{s} } - 1 \right) ; \\
\label{EWeq:K} K & = & \frac{\pi}{2} \int_{d}^{\lambda_{\mathrm{B}}}  \dif r \; r^{2} \exp \left( \frac{2 \lambda_{B} }{r} \right),
\end{eqnarray}
where the fitted value for $\rho_{s}$ is used. Here $K$ is the equilibrium constant for the formation of Bjerrum pairs: dipole-like clusters of two oppositely charged ions that are closely bound due to the strong interaction energy~\cite{Valeriani0,Valeriani1}. In Ref.~\citereg{Valeriani1} it was shown that $\alpha \lesssim 0.5$ implies that strong-coupling effects are not relevant, i.e., the lower the value of $\alpha$ the lower the concentration of Bjerrum pairs (for $\alpha = 1$ all ions have formed pairs and higher order clusters).

The local dimensionless charge density for an equivalent, homogeneously charged colloid is $\mathsf{y} \equiv Z \lambda_{\mathrm{B}}/(\kappa a^{2}) = 4\pi \sigma \lambda_{\mathrm{B}}/\kappa \approx 5.2$ for the parameters of Fig.~\ref{EWfig:dipole_moment}. The parameter $\mathsf{y}$ can be used to estimate the level of nonlinearity in the system and since $\mathsf{y} \approx 5.2$ exceeds unity, mode coupling occurs~\cite{Boon,Boonpatchy}. We will come back to this parameter in the context of heterogeneous surface charges later. For the Janus dipole of Fig.~\ref{EWfig:dipole_moment} we can clearly observe a nonlinear effect, namely $\lim_{r \downarrow a} \rho_{\pm, 0}(r) \ne \rho_{s}$, despite the fact that the charge on the colloid has no intrinsic monopole component. Nonvanishing quadrupole ($\ell = 2$) modes are also induced by mode coupling (nonlinearity)~\cite{Boon}.

In order to quantify the difference between results obtained by MC simulations and by PB theory we compare the difference in the distribution of ions in the double layer directly for each mode. To that end we introduce the so-called difference functions $f_{\ell}$, which can be applied to a general Janus particle with $Q_\mathrm{U}$ unit charges on the \emph{upper} hemisphere and $Q_\mathrm{L}$ unit charges on the \emph{lower} hemisphere, respectively as
\begin{equation}
\label{EWeq:fdef} f_{\ell} = \frac{4\pi}{\vert Q_\mathrm{U} \vert + \vert Q_\mathrm{L} \vert} \int_{a + d/2}^{L/2} \dif r \; r^{2} \left\vert \rho_{\ell,\mathrm{MC}}(r) - \rho_{\ell,\mathrm{PB}}(r) \right\vert ,
\end{equation}
where the ionic charge modes are defined by $\rho_{\ell}(r) = \rho_{+, \ell}(r) - \rho_{-, \ell}(r)$ and where we use the labels MC and PB to indicate the origin of the respective profiles. Equation~\cref{EWeq:fdef} has the property that all $f_{\ell}$ are $0$ when the two profiles are exactly the same and that at least one $f_{\ell} > 0$ when they are not. Because we compare results for the cubic geometry of the simulation box to the spherical geometry of the WS cell in PB theory, the upper integration boundary is set to $L/2 < R$. In principle the difference in shape and associated boundary conditions imply that we compare a simple-cubic crystal of colloids with a liquid of colloids at the same volume fraction. However, due to the separation of the particles and the level of ionic screening the results are virtually independent of the shape of the volume when we compare up to $r = L/2$. This is the reason why we only consider systems with added salt.

In Eq.~\cref{EWeq:fdef} the functions $f_{\ell}$ are `normalized' by $\vert Q_\mathrm{U} \vert + \vert Q_\mathrm{L} \vert$ such that for a homogeneously charged particle $f_{0} = 2$, for the worst-case scenario of full discrepancy between the MC and PB results. Because the counter charge in the double layer should compensate for the net charge on a colloid, each $\vert \rho_{0}(r) \vert$ separately contributes at most $\vert Q_\mathrm{U} \vert + \vert Q_\mathrm{L} \vert$, which explains the normalization. An example of a significant mismatch between the MC and PB results is found deep in the strong-coupling regime where the MC method predicts a total condensation of counter ions in a very small region close to the surface, whilst PB theory predicts that the counter charge is located in a diffuse layer around the colloid of significant width. For higher order modes the value of $f_{\ell}$ is bounded, but the range is not necessarily $[0,2]$. To get a feeling for the order of magnitude of $f_{\ell}$ in the case of a good agreement between PB and MC results, we mention that $f_{0} \approx 0.003$, $f_{1} \approx 0.073$, $f_{2} \approx 0.006$, and $f_{3} \approx 0.057$ for the parameter set shown in Fig.~\ref{EWfig:dipole_moment}, whilst Fig.~\ref{EWfig:fmono}d shows profiles with $f_{0} \approx 0.99$, $f_{1} \approx 0.027$, $f_{2} \approx 0.034$, and $f_{3} \approx 0.031$. Note that although $f_{0}$ in Fig.~\ref{EWfig:fmono}d confirms the huge mismatch between PB and MC results for the $\ell=0$ mode, the $f_{\ell}$ with $\ell > 0$ are still small. This suggests that the spherical symmetry of the problem is only slightly broken within MC simulations and thus that strong coupling does not induce significant multipolar charge distributions inside the screening cloud of a homogeneously charged particle.

\subsection{\label{EWsub:homogen}Homogeneously Charged Spherical Particles}

\begin{figure}
\begin{center}
\includegraphics[width= 8cm]{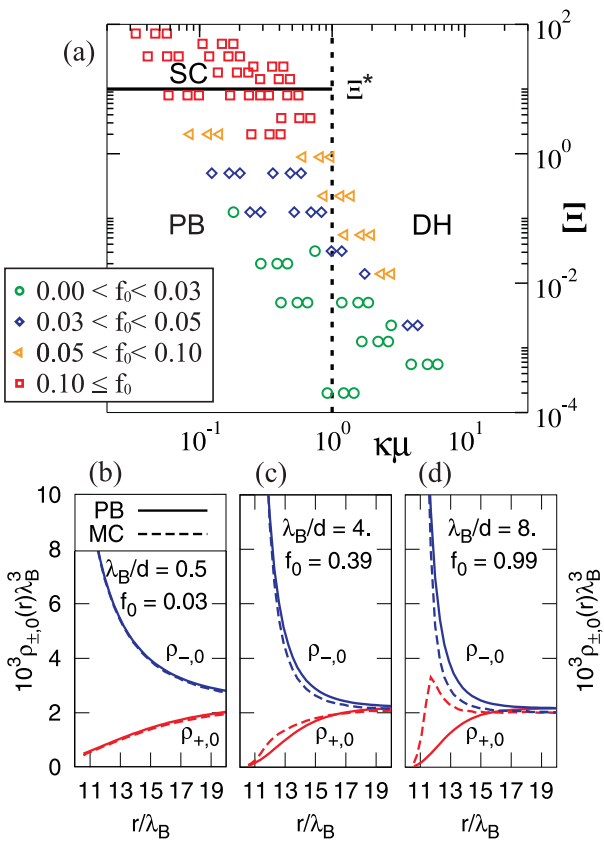}
\end{center}
\caption{\label{EWfig:fmono}A comparison of the data obtained by PB theory and by MC simulations of homogeneously charged spheres according to the difference function $f_{0}$ of Eq.~\cref{EWeq:fdef}, which quantifies the deviation in the distribution of charge in the ionic double layer. We show $f_{0}$ as a function of $\kappa \mu$, the ratio of the Gouy-Chapmann and the Debye length, and $\Xi$, the strong-coupling parameter, for several of the systems we studied. The field-theoretical prediction of Refs.~\citereg{Boroud,Punk} for homogeneously charged flat surfaces partitions parameter space into three regimes, as is indicated by the continuous and the dashed line. The Debye-H\"uckel (DH), Poisson-Boltzmann (PB), and Strong-Coupling (SC) approximations should be used to obtain acceptable results in the respective domains. The value of the PB-SC divide is denoted by $\Xi^{*} \approx 10$. (b-d) Three samples of the ion profiles that are obtained by PB theory (full curves) and MC simulations (dashed curves), showing cation and anion densities. Here $f_{0} = 0.03$, $0.39$, and $0.99$. The discontinuity in the first derivative of the MC $\rho_{+,0}$ profile in (d) is caused by the positive ions condensing on the surface of the colloid in combination with the binning procedure we applied to average the ion densities. A similar discontinuity is present in the $\rho_{-,0}$ results, although this is not visible on the scale of this plot.}
\end{figure}

To prove that the difference functions introduced in Eq.~\cref{EWeq:fdef} give a useful description of the deviation between the MC and PB results, we investigate the monopole deviation $f_{0}$ for homogeneously charged spherical particles, in Fig.~\ref{EWfig:fmono}a. We compare the behaviour of $f_{0}$ to field-theoretical predictions~\cite{Boroud,Punk}, which are also aimed at establishing a range of validity for PB theory, and show that there is a good correspondence between the two ranges.

In Refs.~\citereg{Boroud,Punk} the parameter regimes are investigated, for which various theoretical approximations give trustworthy results for the effective ion distribution of a homogeneously charged flat surface. For these flat surfaces parameter space is partitioned into three pieces, see Fig.~\ref{EWfig:fmono}a. (I) A region where the Debye-H\"uckel (DH) approximation~\cite{Hunter} can be used, the screening is linear. (II) A region where the charge of the surfaces becomes higher, the nonlinear PB equation~\cite{Chapman,Alexander} has to be solved in order to determine the effective electrostatic interactions. (III) A region in which the ion-ion correlations close to the surface require the use of Strong-Coupling (SC) theory~\cite{Rouzina,Grosberg,Moreira}. In Fig.~\ref{EWfig:fmono} the parameters $\kappa \mu = 2/\mathsf{y}$ and $\Xi = (\mathsf{y}/2) \kappa \lambda_{\mathrm{B}}$ represent parameter space in a `field-theoretical language' (see Ref.~\citereg{Punk}), with $\mathsf{y}$ `our' local dimensionless charge density [as defined just below Eq.~\cref{EWeq:K}] and $\mu$ the Gouy-Chapmann length. The Gouy-Chapmann length must not to be confused with the ionic chemical potential, which is also usually denoted by $\mu$. $\Xi$ is a measure for the particle's surface charge, expressed in terms of the Bjerrum length. PB theory produces satisfactory results for $\Xi < \Xi^{*}$, with $\Xi^{*}$ a cut-off value: Ref.~\citereg{Punk} sets the value of $\Xi^{*} \approx 10$ for the transition between the PB and SC regimes. The value of $\kappa \mu$ is of minor importance when $\Xi < \Xi^{*}$ as PB theory can be straightforwardly applied to the DH region in the low-charge limit.

By comparing the MC and PB ion profiles for the $99$ systems containing a homogeneously charged sphere that we have investigated, we found $f_{0}^{*} = 0.1$ to be a good boundary value for the regime ($f_{0} < f_{0}^{*}$) in which PB-theory accurately describes the ion density profiles. Figures~\ref{EWfig:fmono}(b-d) show three example ion profiles to illustrate the possible level of deviation corresponding to a particular $f_{0}$ value: $f_{0} = 0.05$, $0.39$, and $0.99$, respectively. Note that our choice of $f_{0}^{*} = 0.1$ is meant to imply that below this value PB theory gives a good description, however, for $f_{0} > f_{0}^{*}$ PB theory may give a reasonably accurate description (see Fig.~\ref{EWfig:fmono}c), but this is not a given and SC theory may be required to give a good description. Using our $f_{0}^{*}$ values as a function of $\mathsf{y}$ and $\Xi$, see Fig.~\ref{EWfig:fmono}a, we locate the PB-SC divide at $\Xi^{*} \approx 1$. Minor changes in the value of $f_{0}^{*}$ do not significantly change the location of the PB-SC transition in parameter space. However, since what is considered an unacceptable level of the discrepancy between PB and MC results is dependent on the quantities/behaviour we are interested in, there is a degree of arbitrariness to our result. Nevertheless, our approach to this problem and our choice for $f_{0}^{*}$ appears justified since we obtain a similar partitioning of parameter space as was found in Ref.~\citereg{Punk}. This was to be expected for a homogeneously charged sphere, since there is only a geometrical difference with respect to a homogeneously charged plate, which for sufficiently large spheres can be considered small close to the sphere's surface. Our results show that even for relatively small spheres (compared to the size of the ions) there is qualitative agreement.

For completeness we comment on the accuracy of our MC result deep inside the strong-coupling regime. The MC results show that a layer of oppositely charged ions can form on the surface of the charged particle. The interaction between the charges (sites and ions) is such that the free ions effectively condense on the particle, see Refs.~\citereg{Levin6,Belloni,Trizac0,Yu_wu_gao,Denton0,Kanduc} for a more comprehensive account of this phenomenology. The ions in the electrolyte experience similarly strong interactions and form Bjerrum pairs. Since we only consider single particle MC trial moves, the formation of Bjerrum pairs interferes with the exploration of phase space in the strong-coupling limit. The clusters hardly move, because most single particle moves that would break up a cluster are rejected based on the energy difference. This results in an ill-converged ensemble average, when the Bjerrum-pair concentration is high ($\alpha \gtrsim 0.5$)~\cite{Valeriani1}. The problem can be overcome by introducing cluster and association-dissociation moves for the Bjerrum pairs to obtain a more efficient sampling~\cite{Valeriani0,Valeriani1}. However, we do not believe that ion condensation and Bjerrum-pair formation will influence our result with regard to the location of $\Xi^{*}$, since these effects only start to play a role for $\Xi \gg \Xi^{*}$, since then $\alpha > 0.5$.

\subsection{\label{EWsub:janus}Janus-Dipole Charge Distributions}

\subsubsection{\label{EWsubsub:januscompare}Comparison of the Monte Carlo and Poisson-Boltzmann Results}

\begin{figure}
\begin{center}
\includegraphics[width= 8cm]{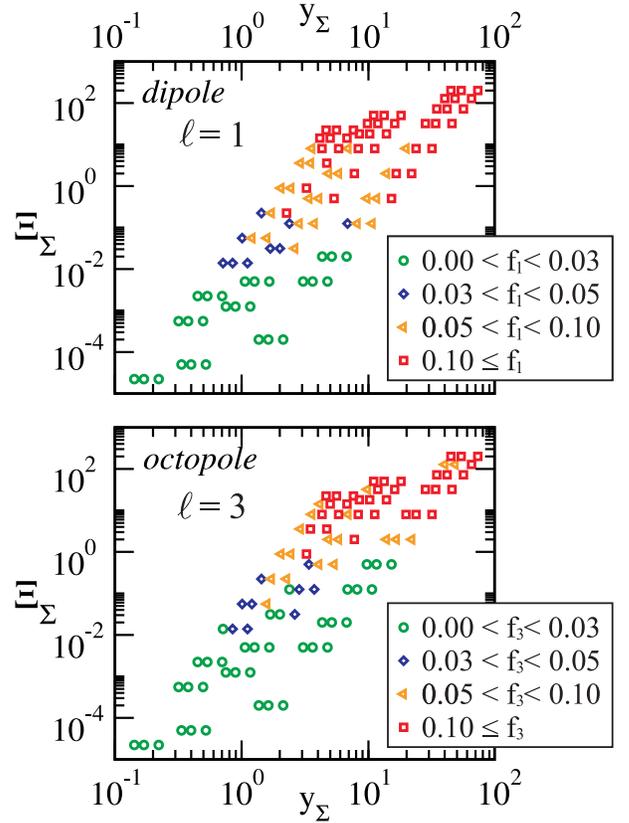}
\end{center}
\caption{\label{EWfig:fdipole}The deviation $f_{\ell}$ in the double layer, determined using the MC simulations and PB theory, for a Janus dipole as a function of the modified charge density $\mathsf{y}_{\Sigma}$ and the modified strong-coupling parameter $\Xi_{\Sigma}$. The subgraphs show the results for the first two odd FL modes ($\ell = 1$, $3$), corresponding to the dipole and octupole term, respectively.}
\end{figure}

In order to investigate the range of applicability of the (nonlinear) Poisson-Boltzmann approximation, we apply our method of comparison from Section~(\ref{EWsub:compare}) to higher order Fourier-Legendre (FL) modes of the ion density for the case of a Janus dipole. Figure~\ref{EWfig:fdipole} shows the deviation parameter $f_{\ell}$ for $\ell=1, 3$ and the large set of parameters we studied. Note that the Janus dipole bears no \emph{even} multipole moments [$\rho_{-,\ell}(r) = \rho_{+,\ell}(r)(-1)^{\ell}$] due to its antisymmetric charge distribution. To apply a representation similar to the one used in Ref.~\citereg{Boroud} for Janus particles, we introduce the following \emph{modified} dimensionless parameters: $\mathsf{y}_{\Sigma} = 2/ (\kappa \mu_{\Sigma}) \equiv  (|Q_{\mathrm{U}}| + |Q_{\mathrm{L}}|) \lambda_{\mathrm{B}} /(\kappa a^2)$ and $\Xi_{\Sigma} \equiv (\mathsf{y}_{\Sigma}/2) \kappa \lambda_{\mathrm{B}} $. The sum of the absolute value of the charge on each hemisphere is used, rather than the total charge (which would be zero in the case of a Janus dipole). For pure monopoles $\mathsf{y}_{\Sigma}$ and $\Xi_{\Sigma}$ reduce to the original parameters $\mathsf{y}$ and $\Xi$. We prefer to express our results in terms of the dimensionless (absolute) local charge density $\mathsf{y}_{\Sigma}$ rather than in terms of $\kappa \mu = 2/\mathsf{y}_{\Sigma}$, since the former is a more natural quantity for PB theory of colloid systems. The use of $\kappa \mu$ in Fig.~\ref{EWfig:fmono} was to illustrate the correspondence between our results and the partitioning given in Ref.~\citereg{Punk}.

For the dipole term ($\ell = 1)$ we observe trends in the value of $f_{1}$ (Fig.~\ref{EWfig:fdipole}) as a function of $\mathsf{y}$ and $\Xi$ similar to those observed for the value of $f_{0}$ of a homogeneously charged sphere (Fig.~\ref{EWfig:fmono}a). The onset of a strong difference in the correspondence between the two results for the dipole mode occurs at $f_{1}^{*} \approx 0.1$. For the octupole ($\ell = 3$) term, the crossover value $f_{3}^{*}$ for an appreciable level of deviation appears to be slightly larger than $0.1$, but on the strength of our results it is difficult to state this with certainty.

Based on Fig.~\ref{EWfig:fdipole}, a regime can be distinguished for the leading dipole term where PB theory yields accurate results for the charge profiles in the electric double layer ($\Xi_{\Sigma} < \Xi_{\Sigma}^{*} \approx 1$). For the $\ell = 1$ through $\ell = 5$ modes ($\ell = 5$ not shown here) the correspondence between MC and PB results is also sufficient when $\Xi_{\Sigma} < \Xi^{*}_{\Sigma}$. The modified parameter $\Xi_{\Sigma}$ therefore appears useful to describe parameter space for dipolar Janus particles with regards to quantifying the region where PB theory can be used to describe the system.

\subsubsection{\label{EWsubsub:janusequations}Multipole-Expanded DLVO Approximation for the Janus Dipole}

Equations describing the electrostatic pair-interaction of (spherical) colloidal particles with a substantial dipolar contribution to the surface charge are derived in the Appendix. This derivation is performed within the Poisson-Boltzmann approximation, and the resulting equations can therefore be considered as an extension of DLVO theory towards inhomogeneously charged particles. The monopole-monopole, monopole-dipole, and dipole-dipole interactions are given respectively by
\begin{subequations}
\label{multipoleinteractions}
\begin{small}
\begin{flalign}
\label{mminteraction} \beta &\Vmmij (\Rijvec) = \lb \frac{\exp(-\kappa R_{ij})}{R_{ij}} Z^{\mathrm{Y}}_i Z^{\mathrm{Y}}_j , \\
\nonumber \beta &\Vmdij (\Rijvec) = \\
\label{mdinteraction} & \lb \frac{\exp (-\kappa R_{ij})}{ R_{ij}^2} \left((\pvec^\mathrm{Y}_i \cdot \bm{\hat{R}}_{ij}) Z^{\mathrm{Y}}_j - (\pvec^\mathrm{Y}_j \cdot \bm{\hat{R}}_{ij}) Z^{\mathrm{Y}}_i\right) , \\
\nonumber \beta & \Vddij (\Rijvec) = \\
\nonumber & \lb \frac{\exp(-\kappa R_{ij})}{R_{ij}^3} \left((1+\kappa R_{ij})(\pvec^\mathrm{Y}_i \cdot \pvec^\mathrm{Y}_j)\right. \\
\label{ddinteraction} & -\left. (3+3 \kappa R_{ij} + (\kappa R_{ij})^2)(\pvec^\mathrm{Y}_i \cdot \bm{\hat{R}}_{ij})(\pvec^\mathrm{Y}_j \cdot \bm{\hat{R}}_{ij}) \right),
\end{flalign}
\end{small}
\end{subequations}
with $\Rijvec$ the distance vector between particle $i$ and $j$, $R_{ij} \equiv \vert \Rijvec \vert$, and $\bm{\hat{R}}_{ij} = \Rijvec / R_{ij}$. We also introduce the `Yukawa-monopoles' $Z^\mathrm{Y}_i$ and `Yukawa-dipoles' $\pvec^\mathrm{Y}_i$, which, for Janus spheres of radius $a$, are given by
\begin{subequations}
\label{monodipoleprefactors}
\begin{eqnarray}
\label{ZZyukawa} Z^{\mathrm{Y}}_i &=& (Q_\mathrm{U}+Q_\mathrm{L}) \frac{\exp (\kappa a)}{1+\kappa a},\\
\nonumber \pvec^\mathrm{Y}_i &=& (Q_\mathrm{U} - Q_\mathrm{L}) \frac{ 3 a\exp (\kappa a) \nvec_{i} }{4 (2 + \epsilon_\mathrm{c}/\epsilon) (1+\kappa a) + (\kappa a)^2} , \\
\label{pyukawa} &~&
\end{eqnarray}
\end{subequations}
with $\nvec_{i}$ is the particle's symmetry axis, which points to the northern hemisphere, $Q_\mathrm{U}$ the total charge on the upper hemisphere, $Q_\mathrm{L}$ the total charge on the lower hemisphere, and $\epsilon_\mathrm{c}/\epsilon$ the ratio between the relative dielectric constant of the colloidal particle and that of the surrounding medium, which we choose $1$ unless stated differently. However, Eqs.~\cref{multipoleinteractions} and~\cref{monodipoleprefactors} are, as is typical for DLVO theory, only valid for sufficiently small charges, since linearised PB theory (Debye-H\"uckel approximation) was employed in the derivation of these equations. Fortunately, charge renormalisation has proven to be a useful tool in broadening the range of applicability towards particles with a higher charge~\cite{Alexander, Trizac1}. This renormalisation procedure was extended towards dipoles and higher multipoles in Ref.~\citereg{Boon}.

\subsection{\label{EWsub:hemi}Hemispherical Charge Distributions}

\subsubsection{\label{EWsubsub:hemicompare}Comparison of the Monte Carlo and Poisson-Boltzmann Results}

For the hemispherical charge distribution, we also compared PB results with those of MC simulations, in order to determine the regime of validity of the PB-multipole expansion. We again considered $99$ systems and found that for the even modes the level of deviation $f_{\ell}^{*} \approx 0.1$ sets a rough upper bound to the applicability of PB theory. For the odd modes the correspondence between the MC and PB results seems to hold for slightly higher values of the deviation parameter: $f^{*}_{\ell} \approx 0.25$. Using these two values of $f^{*}$ we can roughly locate the range of validity of the PB result in the region $\Xi_{\Sigma} \lesssim 1$. The effects of strong coupling are however far more apparent for our hemispherical charge distribution than for the other two distributions we considered. This is because in the MC simulations we used $50$ divalent charge sites on the upper hemisphere instead of monovalent sites. Our results for the hemisphere are therefore less convincing than for the other two charge distributions, but our preliminary indication is that the range in which the PB approximation is valid is roughly the same as that for the monopole and Janus dipole.

\subsubsection{\label{EWsubsub:hemiequations}Multipole-Expanded DLVO Approximation for the Hemispherical Charge Distribution}

As mentioned in the previous section, one cannot always resort to standard DLVO theory for particles with an anisotropic surface charge, because dipole, quadrupole, or higher order multipole charge distributions are relevant. Consequently, two-body interactions of the form monopole-dipole, dipole-dipole, monopole-quadrupole, etc., should be considered as well.

\begin{figure}[!ht]
\begin{center}
\includegraphics[width= 8cm]{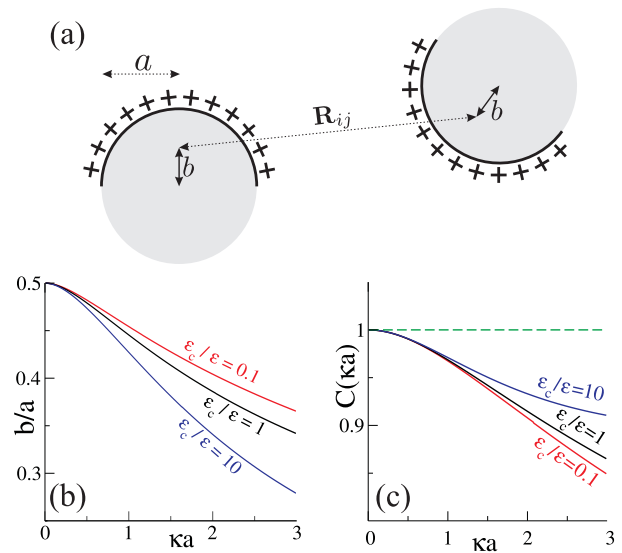}
\end{center}
\caption{\label{fig:chargedistribution}(a) A sketch of two interacting particles with radius $a$, both having charge on only one hemisphere. To eliminate the dipole moment we relocate the centre of the charge distribution at a distance $b$ (along the particle's rotational symmetry axis) from the geometrical (hard-core) centre of the particle. This choice results in a distance $R_{ij}$ between the charge distributions. Graphs (b) and (c) show the suggested value of $b$ and the Yukawa weight factor $C(\kappa a)$, respectively, as a function of the colloid radius $a$ in terms of the Debye screening length $\kappa^{-1}$, for a wide range in $\epsilon_{\mathrm{c}} / \epsilon$, the ratio between the relative dielectric constant of the particle and that of the surrounding medium. In (c), we indicate $C = 1$, which is the weight in case of the regular DLVO equation, using a dashed line.}
\end{figure}

Figure~\ref{fig:chargedistribution}a shows two spherical particles with a positive hemispherical charge distribution, i.e., one side is charged the other is uncharged. Instead of choosing the centre of the charge distribution to coincide with the geometrical centre of a spherical particle one is free to place this point anywhere inside the particle. The most natural location is the point for which the (Yukawa-)dipole moment vanishes and thereby all dipole interactions. With this choice the electrostatic two-body interaction in terms of only a monopole-monopole term, is expected to be maximally accurate. This point is located on the rotational symmetry axis of the hemispherically charged colloid. For sufficiently large interparticle distances we obtain the following interaction potential for the shifted-monopole approximation:
\begin{equation}
\label{eq:mono-mono}\beta \Vij (R_{ij}) \approx \left( Z \frac{e^{\kappa a}}{1 + \kappa a} C (\kappa a) \right)^2 \lb \frac{\exp(-\kappa R_{ij})}{R_{ij}},
\end{equation}
with $R_{ij}$ the shifted centre-to-centre distance, $Z$ the particle charge, and $C(\kappa a)$ a renormalisation factor. This renormalisation factor depends on the ratio of the particle radius $a$ and the Debye screening length $\kappa^{-1}$ only and $C = 1$ in case of regular DLVO theory. We can numerically determine the distance $b$ from the particle's geometrical centre, where we should place the shifted monopole such that the dipole term vanishes. Figure~\ref{fig:chargedistribution}b shows the ratio $b/a$ as a function $\kappa a$ for several values of $\epsilon_{\mathrm{c}}/\epsilon$. Note that $b/a$ goes to $1/2$ for small $\kappa a$, which is the (Coulomb) result that is obtained for unscreened particles. We show the corresponding renormalisation factors $C(\kappa a)$ in Fig.~\ref{fig:chargedistribution}(c). We find $C(\kappa a) \lesssim 1$, because the charge is located closer to the (new) centre of the charge distribution than for homogeneously charged particles. Note that the interaction potential is independent of the orientation of the particles, when rotated around the shifted (monopole-charge) centre. The monopole-monopole approximation therefore does not capture the full interaction between the two hemispherically charged colloids. To capture the orientational dependence higher order modes (octopole and higher) are required. However, for a monopole-monopole only approximation Eq.~\cref{eq:mono-mono} is maximally accurate by design. In the Appendix we explain how to calculate and, thereby, also how to minimize the Yukawa dipole moment for any type of charge distribution and hard-core shape.

\section{\label{EWsec:disc}Conclusion and Outlook}

In this paper we studied the range in parameter space for which the nonlinear Poisson-Boltzmann (PB) theory accurately describes the behaviour of the ions around a Janus charge-patterned spherical colloid in a 1:1 electrolyte. We used primitive-model Monte Carlo (MC) simulations to establish the ion density around such a charged particle for a huge set of parameters. By also computing the ion density for the same parameters and comparing the two results, we were able to establish a regime in which this PB theory gives a good approximation for the ion distribution. This comparison is based on Fourier-Legendre (FL) decomposition of the MC ion density to determine the contribution of the monopole, dipole, quadrupole, $\dots$ charge terms. The theoretical approach also relies on FL decomposition and this enables us to quantify the differences on a mode-by-mode basis.

For a homogeneously charged sphere we compared our range of validity for PB theory to the range found in Refs.~\citereg{Boroud,Punk} for a system of homogeneously charged flat plates in an electrolyte. There is a remarkable correspondence between the two ranges, especially considering the small size of the colloids that we studied in relation to the size of the ions. For such small spheres a greater deviation with respect to the results of a flat-plate calculation could reasonably be expected. We were also able to show that the range in which the PB results accurately describe the ion density around a spherical Janus-dipole is similar to that found for the homogeneously charged sphere. For particles with only one (homogeneously) charged hemisphere, there is an indication that the regime in which PB theory can be applied matches the regime found for the two other particles.

In the PB-regime that we obtained, we can use simple (multipole-expanded) DLVO-like equations, which we derived in this paper, to describe the interactions between two particles with a Janus-type charge. We gave explicit expressions for the monopole and dipole interactions, since these terms are typically dominant for Janus particles. These electrostatic interactions resemble well-known Yukawa interactions, and reduce to these in the homogeneous charge limit. For the Janus-dipole obtaining the (multipole-expanded) DLVO expression is relatively simple. To accurately model a hemispherical charge distribution using only a monopole-term is a little more complicated. The key step proved to be shifting the centre of the charge distribution from the centre of the particle towards the charged hemisphere in order to eliminate the Yukawa-dipole contribution.

Our analysis forms a basis of a good understanding of the range in parameter space for which the PB approximation can be applied to describe the behaviour of heterogeneously charged colloids. This is, for instance, relevant to the study of such particles using simulations, where PB-theory-based effective interactions can be used to study the phase behaviour of such particles in the right regime. Note that we only considered equilibrium ion density profiles of stationary colloids. The rotational movement of mobile charge-patterned colloids can occur on time scales that would lead to an out-of-equilibrium double layer. What effect the out-of-equilibrium ion density would have on the screening of the particle and how such effects should be incorporated into effective interaction potentials used in simulations, is left for future investigation.

\section{\label{sec:ackn}Acknowledgements}

M.D. acknowledges financial support by a ``Nederlandse Organisatie voor Wetenschappelijk Onderzoek'' (NWO) Vici Grant and R.v.R. by the Utrecht University High Potential Programme and by a NWO ECHO grant.

\begin{appendix}

\section*{\label{EWsec:Janus_Debye}Appendix: Janus Particles and the Debye-H\"uckel Approximation - DLVO Theory for Patchy Colloids}

In this appendix we derive the (multipole-expanded) DLVO-like equations for general charge distributions and we show how this leads to Eqs.~\cref{multipoleinteractions}~-~\cref{monodipoleprefactors}. Before that we set the stage by first examining the quality of the dipole-only approximation for a Janus-type charge distribution.

\begin{figure}[!ht]
\begin{center}
\includegraphics[width = 8cm]{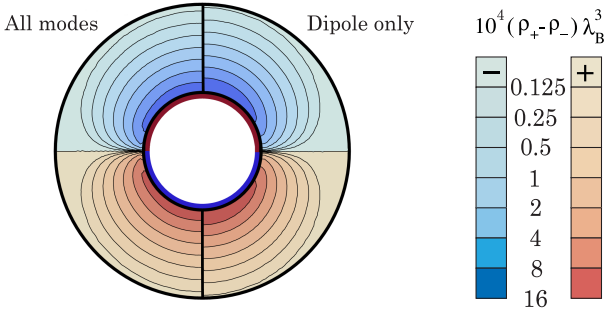}
\end{center}
\caption{\label{fig:pbdipolecompare} A contour plot of the net charge density $\rho_+(\rvec) - \rho_-(\rvec)$ around an antisymmetric Janus particle for the parameters $Z=10$, $N_{\pm} = 425$, $\lb = \iondiam$, and $a= 10 \lb$, showing the profile that follows from a mode expansion up to $\ell=6$ on the left, on the right only the dipole mode is plotted.}
\end{figure}

We have shown that the mode expansion up to $\ell=6$ gives good agreement between MC simulations and PB theory regarding the ion profiles in a well-defined regime of parameters. For Janus particles in general the most dominant multipoles are the monopole and/or the dipole; note, however, that for `pure' Janus dipoles the monopole term vanishes. To show this Fig.~\ref{fig:pbdipolecompare} plots the PB result for the local ionic charge densities around such a purely dipolar Janus particle, using a smaller charge ($Z=10$) than elsewhere in this paper. The number of ions added to the system, $N_{\pm} = 425$, corresponds to $\kappa a \approx 2.8$. We find that $\mathsf{y}_{\Sigma} = 0.36$ and $\Xi_{\Sigma} = 0.05$ for this system, which implies that we are within the regime where PB-theory is applicable according to our analysis. We are aware that a multipole expansion of Yukawa-like interaction potentials does not necessarily converge in general~\cite{Agra0,Agra1}. In this particular case, however, we catch most of the physics of the interacting Janus particles, by treating the particle as a `pure' dipole (without higher order modes). To illustrate this, the left side of Fig.~\ref{fig:pbdipolecompare} shows the ion charge density up to $\ell=6$, which we proved to be sufficient for good correspondence with MC results in the regime where PB-theory is applicable. The right side shows the dipole $(\ell = 1)$ mode only. The differences are therefore due to the missing $\ell=3$ and $\ell=5$ terms. The dipole approximation overestimates the electrostatic potential in the axial direction, whilst it underestimates the electrostatic potential in the perpendicular direction.

With this in mind we explain the way to extend the applicability of DLVO theory to anisotropically charged particles. The result of setting up this theory allows us us to find explicit equations for the monopole-dipole and dipole-dipole interaction potential between Janus particles as a function of their orientation. These expression are similar to the well-known expressions for the interaction of unscreened dipoles. We begin by considering the effective electrostatic energy of an extended charge configuration $e q(\rvec)$ in a 1:1 electrolyte with bulk concentration $2\rho_{s}$ - we do not incorporate hard-core effects at present. By using the electrostatic energy from Coulomb's law combined with the ideal-gas entropy for the monovalent ions, we find that the grand potential of the charge configuration inside a two-component monovalent ion mixture inside a solvent is given by
\begin{flalign}
\label{linearhamiltoniandipole} \beta H &= \int\dif\rvec~ \sum_{\alpha = \pm} \rho_\pm(\rvec) \left(\frac{\rho_\pm
(\rvec)}{\rho_s} - 2\right)\nonumber\\
& + \frac{1}{2}\int\dif\rvec~ (\rho_+(\rvec) - \rho_-(\rvec)+ q(\rvec))\phi(\rvec),
\end{flalign}
where $\rho_\pm(\rvec)$ are the ion densities and where the dimensionless electrostatic potential is
\begin{equation}
\label{phidefcd} \phi(\rvec) = \lambda_{\mathrm{B}} \int\dif\rvec' \frac{\rho_+(\rvec') - \rho_-(\rvec') + q(\rvec')}{|\rvec - \rvec'|} .
\end{equation}
Note that we applied the Debye-H\"uckel approximation; the `usual' entropic term of the ions is linearised via $\rho_\pm(\rvec) (\log (\rho_\pm(\rvec)/\rho_s) - 1) \approx \rho_\pm(\rvec) (\rho_\pm(\rvec)/\rho_s - 2)$, implying that we assume the ion densities not to vary much from the bulk value $\rho_s$. For a more detailed derivation of Eq.~(\ref{linearhamiltoniandipole}) see for example Ref.~\citereg{boon_phd}.

\begin{figure}[!ht]
\begin{center}
\includegraphics[width = 8cm]{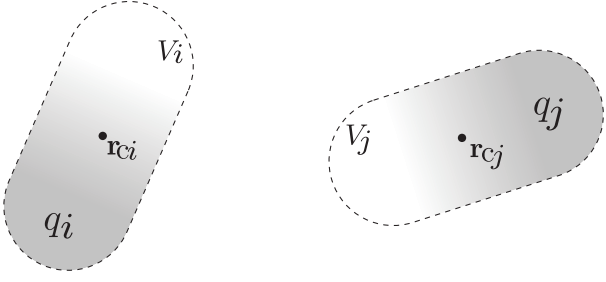}
\end{center}
\caption{\label{fig:appendixintro} A sketch of two interacting charge distributions (grey) $q_i(\rvec)$ and $q_j(\rvec)$, inside the enclosing volumes $V_i$ and $V_j$, respectively. These volumes are rod-like in this particular example to resemble rods with one charged `head' and have boundaries that are indicated by dashed instead of solid lines since we do not consider hard cores while calculating the ion densities connected to these charge distributions.  The centres of the charge distributions can be chosen arbitrarily and are indicated by $\rvec_i$ and $\rvec_j$.}
\end{figure}

We consider a system that consists of a collection of $M$ (fixed) charges located at $\rvec_{\mathrm{c} i}$, with $i = 1$, $\dots$, $M$ an index. These charges should not be considered as point charges, but rather as localized charge distributions $q_i(\rvec)$ inside associated volumes $V_i$, which are non overlapping and centred at $\rvec_{\mathrm{c} i}$, as is sketched in Fig.~\ref{fig:appendixintro}. Each $q_i(\rvec)$ is only nonzero inside $V_i$, and the total (non-ionic) charge density is hence written as $q(\rvec) = \sum_{i=1}^{M} q_i(\rvec)$.

Since we have not included any hard-cores yet, the ion densities follow immediately from setting the functional derivative of Eq.~(\ref{linearhamiltoniandipole}) w.r.t. the ion densities to zero, $\delta H / \delta \rho_\pm (\rvec)= 0$, yielding  $\rho_{\pm}(\rvec) = \rho_s (1 \mp \sum_{i=1}^{M} \phi_i(\rvec))$, with
\begin{eqnarray}
\label{eq:phirvec} \phi_i(\rvec) &=& \lambda_{\mathrm{B}} \int_{V_i}\dif\rvec'~q_i(\rvec')\frac{\exp (-\kappa |\rvec' - \rvec|)}{|\rvec' - \rvec|},
\end{eqnarray}
where $\kappa^{2} = 8 \pi \lambda_{B} \rho_{s}$. By assuming that the charges $q_i(\rvec)$ within the volumes $V_i$ have a fixed position, Eq.~(\ref{linearhamiltoniandipole}) can be written, up to a self-energy constant, as
\begin{equation}
\label{dipoleinteractionenergy} \beta H = \lambda_{\mathrm{B}} \sum_{i<j} \iint_{V_i,V_j}\dif\rvec~\dif\rvec{'}~ q_i(\rvec) q_j(\rvec') \frac{\exp{(-\kappa |\rvec' - \rvec|)}}{|\rvec' - \rvec|} .
\end{equation}
Equations~\cref{eq:phirvec}~and~\cref{dipoleinteractionenergy} were derived using the Taylor-expanded (quadratic) Hamiltonian~(\ref{linearhamiltoniandipole}) and therefore give appropriate results only in case the dimensionless electrostatic potential $\phi(\rvec)$ remains sufficiently small w.r.t. unity.

We will now use the result of Eq.~(\ref{dipoleinteractionenergy}) to obtain an analogous theory for charge distributions \emph{with} associated hard-core volumes. This is done by `freezing' the \emph{ionic} charge profiles inside the volumes $V_i$ and adding these to the charge configurations $q_i(\rvec)$, such that the new charge distributions $\tilde{q}_i(\rvec)$ are found. Effectively we compensate for the fact that ions in the Yukawa approximation can penetrate the hard particle. We thus consider a new combined charge distribution $\tilde{q}_i(\rvec)$ that is exactly the distribution we are interested in, by compensating the fixed charge for the ion profiles it induces.

Starting with the obtained ion densities for the system without hard cores, we split the entire system volume $\V$ into $\Vion$, consisting of all points $\rvec$ outside the volumes $V_i$ for all $i$, and the volume $\Vcol$ of points that are inside one of these volumes. Note that $\Vcol$ is the complement of $\Vion$. The ion densities are also split into $\rho_{\pm}^{\mathrm{out}}(\rvec)$ and $\rho_{\pm}^{\mathrm{in}}(\rvec)$ such that $\rho_{\pm}^{\mathrm{out}}(\rvec)+\rho_{\pm}^{\mathrm{in}}(\rvec) = \rho_\pm(\rvec)$ and $\rho_{\pm}^{\mathrm{out}}(\rvec)=0$ for all $\rvec$ inside $\Vcol$, and $\rho_{\pm}^{\mathrm{in}}(\rvec)=0$ for all $\rvec$ inside $\Vion$. Eq~(\ref{linearhamiltoniandipole}) may therefore be rewritten as
\begin{flalign}
\nonumber \beta H = \int_{\Vion} &\dif\rvec~ \sum_{\alpha = \pm} \rho_{\pm}^{\mathrm{out}}(\rvec) \left(\frac{\rho_{\pm}^{\mathrm{out}}
(\rvec)}{\rho_s} - 2\right) \\
\nonumber + \int_{\Vcol} &\dif\rvec~ \sum_{\alpha = \pm} \rho_{\pm}^{\mathrm{in}}(\rvec) \left(\frac{\rho_{\pm}^{\mathrm{in}}
(\rvec)}{\rho_s} - 2\right) \\
\label{splithamiltoniandipole} +\int_{\phantom{\Vcol}} &\dif\rvec~ \frac{1}{2}(\rho_{+}^{\mathrm{out}}(\rvec) - \rho_{-}^{\mathrm{out}}(\rvec) + \tilde{q}(\rvec))\phi(\rvec) ,
\end{flalign}
with
\begin{equation}
\label{phidefhc} \phi(\rvec) = \lambda_{\mathrm{B}} \int\dif\rvec' \frac{\rho_{+}^{\mathrm{out}}(\rvec')-\rho_{-}^{\mathrm{out}}(\rvec')+\tilde{q}(\rvec')}{|\rvec - \rvec'|} ,
\end{equation}
and the `new' colloidal charge distributions
\begin{eqnarray}
\nonumber \tilde{q}(\rvec) &=& q(\rvec) + \rho_{+}^{\mathrm{in}}(\rvec) - \rho_{-}^{\mathrm{in}}(\rvec) \\
\label{Q_rvec} &=& \sum_{i=1}^{M} \tilde{q}_i(\rvec),
\end{eqnarray}
for which $\tilde{q}_i(\rvec) \equiv q_i(\rvec) + \rho_{+}^{\mathrm{in}}(\rvec) - \rho_{-}^{\mathrm{in}}(\rvec)$ when $\rvec$ inside $V_i$.

Within the DLVO-approximation, the net ionic charge density $\rho_{+}^{\mathrm{in}}(\rvec) - \rho_{-}^{\mathrm{in}}(\rvec) $ in volume $V_i$ is only induced by the fixed charges $q_i(\rvec)$ in that volume itself (large inter-particle distances). One therefore finds for $\rvec$ in $V_i$ that
\begin{flalign}
\nonumber \tilde{q}_i(\rvec) &\approx q_i(\rvec) -2\rho_s\phi_i(\rvec) \\
\label{HCchargeequation} &= q_i(\rvec) -2\lb\rho_s\int_{V_i}\dif\rvec' ~q_i(\rvec') \frac{\exp (-\kappa |\rvec' - \rvec|)}{|\rvec' - \rvec|} .
\end{flalign}
The second line in Eq.~(\ref{splithamiltoniandipole}) becomes a constant that may regarded as a self-energy term and therefore can be ignored to yield
\begin{flalign}
\nonumber \beta H &= \int_{\Vion} \dif\rvec~ \sum_{\alpha = \pm} \rho_\pm(\rvec) \left(\frac{\rho_\pm
(\rvec)}{\rho_s} - 2\right) \\
\label{linearhamiltoniandipole2} &+ \frac{1}{2}\int\dif\rvec~ (\rho_+(\rvec) - \rho_-(\rvec) + \tilde{q}(\rvec))\phi(\rvec) ,
\end{flalign}
with the restriction that $\rho_\pm(\rvec)=0$ if $\rvec$ is inside $\Vcol$. Comparing Eq.~\cref{linearhamiltoniandipole} with Eq.~\cref{linearhamiltoniandipole2}, the effective interaction Hamiltonian between hard particles with charge densities $\tilde{q}_i(\rvec)$ can be recognized in the latter. The effective interaction energy for such a system can thus be obtained from Eq.~(\ref{dipoleinteractionenergy}) with $q_{i}(\mathbf{r})$ the solution of Eq.~\cref{HCchargeequation}, with $\tilde{q}_{i}(\mathbf{r})$ the actual charge density of interest. For particles with hard-core volume $V_i$ and corresponding charge densities $\tilde{q}_i(\rvec)$ it is necessary to first solve $q_i(\rvec)$ from Eq.~(\ref{HCchargeequation}). By finding $q_{i}(\rvec)$ we obtain the charge distribution that compensates for the self-induced ion densities, since these are required in to determine the effective interaction in Eq.~(\ref{dipoleinteractionenergy}). 

\begin{figure}[!ht]
\begin{center}
\includegraphics[width = 8.0cm]{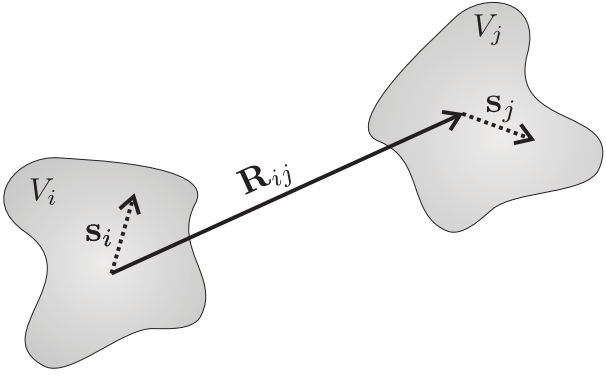}
\end{center}
\caption{\label{fig:chargedistribution_sketch} A sketch of two interacting charge distributions $q_i$ and $q_j$ with hard-core volumes $V_{i}$ and $V_{j}$, respectively, separated by a distance $\Rijvec$. We also show the vectors $\svec_i$ and $\svec_j$, which have their origins at $\rvec_{\mathrm{c} i}$ and $\rvec_{\mathrm{c} j}$, respectively.}
\end{figure}

At this point we have found a procedure to calculate the electrostatic interactions between inhomogeneously charged colloidal particles with hard cores, within the DLVO approximation that we discussed earlier. In order to isolate specific multipole interactions between colloids, Eq.~(\ref{dipoleinteractionenergy}) is rewritten as
\begin{flalign}
\nonumber \beta H = \lambda_{\mathrm{B}} \sum_{i<j} \iint_{V_i, V,j}&\dif\svec_i\dif\svec_j~ \left\{\vphantom{\frac{2}{2}}q_i(\rvec_{\mathrm{c} i}+\svec_i) q_j(\rvec_{\mathrm{c} j}+\svec_j)\right. \\
\label{linearchargedistributionhamiltonian} & \left.\cdot \frac{\exp (-\kappa |\Rijvec + \svec_j - \svec_i|)}{|\Rijvec +\svec_j - \svec_i|}\right\} ,
\end{flalign}
with $\Rijvec = \rvec_{\mathrm{c} j}-\rvec_{\mathrm{c} i}$ the colloid-colloid distance and $\svec_i = \rvec - \rvec_{\mathrm{c} i}$ the coordinate relative to $\rvec_{\mathrm{c} i}$. Both are shown in Fig.~\ref{fig:chargedistribution_sketch}. The Yukawa interaction can then be expanded into spherical harmonics around these centres~\cite{Nozawa}, which are $\rvec_{\mathrm{c} i}$ and $\rvec_{\mathrm{c} j}$. For this we must assume that the distance from any point in $V_i$ to its centre $\rvec_{\mathrm{c} i}$ is less than the distance to any other centre $\rvec_{\mathrm{c} j}$. This however automatically holds for equi-sized spherical volumes if one chooses the centres $\rvec_{\mathrm{c} i}$ exactly in the middle of the spheres.

Here, the monopole and dipole terms are of main interest. Eq.~(\ref{linearchargedistributionhamiltonian}) can therefore be written, up to a monopole-monopole, monopole-dipole, and dipole-dipole interaction term as
\begin{equation}
\label{multihamiltonian} H = \sum_{i<j} \Vmmij (R_{ij}) + \Vmdij (\Rijvec) + \Vddij (\Rijvec) .
\end{equation}
The resulting interaction terms are given by Eq.~(\ref{multipoleinteractions}), in which the `Yukawa monopole' and `Yukawa dipole' are now (for the general charge distribution)
\begin{subequations}
\begin{eqnarray}
\label{ziintegral} Z^\mathrm{Y}_i &=& \int_{V_i}\dif \svec_i~q_i(\rvec_{\mathrm{c} i}+\svec_i) \frac{\sinh \kappa s_i}{\kappa s_i}; \\
\nonumber \pvec^\mathrm{Y}_i &=& 3\int_{V_i}\dif \svec_i~q_i(\rvec_{\mathrm{c} i}+\svec_i)\left(\frac{\cosh \kappa s_i - \frac{1}{\kappa s_i}\sinh \kappa s_i}{(\kappa s_i)^2}\right) \svec_i , \\
\label{yukawadipole} & &
\end{eqnarray}
\end{subequations}
with $s_{i} = \vert \mathbf{s}_{i} \vert$. Note that Eqs.~(\ref{ziintegral}) and~(\ref{yukawadipole}) reduce to the well-known expressions for Coulomb monopole and dipole in the limit $\kappa a \downarrow 0$, see Ref.~\citereg{Griffiths}. Also the interaction terms (\ref{multipoleinteractions}) give the well-known result for pure Coulomb systems in this limit. Finally, the electrostatic potential around charge distribution $i$ can be expanded as
\begin{eqnarray}
\phi_i(\rvec_{\mathrm{c} i} + \svec_i) &=& Z^{\mathrm{Y}}_i \lb \frac{\exp(-\kappa s_i)}{s_i}\nonumber\\
\nonumber &+& (\pvec^\mathrm{Y}_i\cdot\hat{\svec}_i) \lb \frac{\exp(-\kappa s_i)}{s_i^2}(1+\kappa s_i) \\
\label{phimonopoledipole} &+& \mathcal{O} (\ell\geq2) ,
\end{eqnarray}
in which quadrupole and higher order multipoles are not included.

As an example we show that the interaction between particles with a homogeneous charge distribution is in agreement with DLVO theory. By considering a spherical colloid with a hard core radius $a$ and a charge $Z$ distributed homogeneously over its surface and choosing $\rvec_{\mathrm{c} i}=\mathbf{0}$ for convenience, the charge distribution is given by $\tilde{q}_i = Z/(4\pi a^2) \delta(s_i - a)$. It can be shown that Eq.~\cref{HCchargeequation} is solved by
\begin{equation}
\label{jDLVO:hc_colloid_qi} q_i(\svec_i) = 
\begin{cases}
\frac{Z}{4\pi a^2} \delta(s_i - a) + \frac{Z \lb 2 \rho_s }{a (1+\kappa a)} & \text{if $s_i \leq a$,} \\
0 &\text{if $s_i>a$.}
\end{cases}
\end{equation}
Note that the `added' charge density inside the colloids has the same sign as the surface charge and exactly cancels the ionic charge due screening. From Eq.~(\ref{ziintegral}) one obtains $Z^{\mathrm{Y}}_i = Z \exp (\kappa a) / (1+\kappa a)$ and this gives the DLVO result as Eq.~(\ref{mminteraction}) shows.

The Yukawa monopole and the Yukawa dipole can also be extracted from the solution for the electrostatic potential outside a spherical particle, using Eq.~(\ref{phimonopoledipole}), \emph{without} calculating the charge density $q_i(\rvec)$ explicitly. Namely, we can often solve the (linearized) Poisson-Boltzmann equation around a single particle mode-by-mode, and then read the multipole moments from the mode amplitudes in the final solution. As an example we consider spherical particles with a fixed surface charge distribution that is rotationally symmetric around the unit vector $\nvec_i$, e.g., a Janus particle. The electrostatic potential in the vicinity of the single particle can be expanded as $\phi(s_i,x_i)=\sum_{\ell=0}^{\infty}\phi_{\ell}(s_i)P_{\ell}(x_i)$, with $x_i=\hat{\svec_i} \cdot \nvec_i$, and $P_{\ell}(x_i)$ the $\ell$'th order Legendre polynomial~\cite{boon_phd}. Now $\ell=0$ and $\ell=1$ correspond to the monopole and the dipole contribution to the electrostatic potential, respectively. The multipole mode functions, which solve the linearised PB equation, behave as $\phi_{\ell}(s_i)\sim s_i^{\ell}$ for $r<a$ and $\phi_{\ell}(s_i)\sim k_i(\kappa s_i)$ for $r>a$, with $k_i$ the $i$'th modified spherical Bessel function. By applying the boundary condition
\begin{equation}
\label{janus_gauss} \lim_{s_i \downarrow a}\phi'(s_i,x_i) = \lim_{s_i \uparrow a} \phi'(s_i,x_i) - 4\pi\lb\sigma(x_i) ,
\end{equation}
with the prime ($'$) denoting the radial derivative (w.r.t. $s_i$), and $\sigma(x_i)$ the surface-charge density, a solution to the electrostatic potential in terms of a multipole expansion is obtained. Since the modified spherical Bessel functions can also be recognized in Eq.~(\ref{phimonopoledipole}), one is able to derive that the monopole and the dipole moment should relate to the expansion of the surface-charge density, $\sigma(x_i)=\sum_{\ell=0}^{\infty}\sigma_{\ell}P_{\ell}(x_i)$, by
\begin{eqnarray}
\label{janusyukawamonopole} Z^{\mathrm{Y}}_i &=& 4\pi a^2 \sigma_0 \frac{\exp (\kappa a)}{1+\kappa a} ; \\
\label{janusyukawadipole} \pvec^\mathrm{Y}_i &=& 4 \pi a^3 \sigma_1 \frac{\exp (\kappa a)   }{3 + 3\kappa a + (\kappa a)^2} \nvec_i .
\end{eqnarray}
In the case of Janus particles with charge densities $\sigma_\mathrm{U} \equiv Q_\mathrm{U} / (4\pi a^2)$ and $\sigma_\mathrm{L} \equiv Q_\mathrm{L} / (4\pi a^2) $ on the upper and lower hemisphere respectively, one finds $\sigma_0 = \frac{1}{2}(\sigma_\mathrm{U} + \sigma_\mathrm{L})$ and $\sigma_1 = \frac{3}{4} (\sigma_\mathrm{U} - \sigma_\mathrm{L})$.

Throughout this paper we assumed that the dielectric constant of the particles matched the dielectric constant of the solvent. However, Eq.~(\ref{janusyukawadipole}) can easily be extended to particles that have a dielectric mismatch with the solvent. We modify Eq.~(\ref{janusyukawadipole}) by including a dielectric jump in the boundary condition [Eq.~(\ref{janus_gauss})], such that the Yukawa multipoles for these particles can be obtained as well. As is known from the DLVO equation, we find that the Yukawa monopole is unaffected by the value of the dielectric constant inside the particle, whilst the Yukawa dipole changes. Equation~(\ref{janusyukawadipole}) becomes
\begin{equation}
\label{janusyukawadipolemismatch} \pvec^\mathrm{Y}_i =  4 \pi a^3 \sigma_1 \frac{\exp (\kappa a)}{(2 + \epsilon_\mathrm{c}/\epsilon) (1+\kappa a) + (\kappa a)^2} \nvec_i,
\end{equation}
with $\epsilon_\mathrm{c}/\epsilon$ the ratio of the relative dielectric constant inside and outside the particles. Note that the dipole moment becomes very small if the interior of the colloid is very polar w.r.t. the surrounding medium, e.g., (Pickering) emulsions of water in oil. Consequently the monopole-dipole and dipole-dipole interactions will be negligible for these systems, even in case of an asymmetric charge distribution. However, if the dielectric constant of the colloid is comparable or smaller than the dielectric constant of the solvent a significant dipolar interaction may arise.

\section*{\label{EWsec:Contr}Contributions}

J.d.G performed the MC simulation studies, N.B. performed the PB-theory calculations, and they collaborated fully on the analysis of their findings and on the preparation of the manuscript; both contributing equally. M.D. and R.v.R. primarily supervised J.d.G. and N.B., respectively. All authors discussed the results and commented on the manuscript. 

\end{appendix}

\end{document}